\documentclass[dvipsnames, 11pt]{scrartcl}

\usepackage{pgfplots}
\usepackage{subcaption}
\usepackage[english]{babel} 
\usepackage{amsmath}
\usepackage{mathtools}
\usepackage{multirow}
\usepackage{tablefootnote}
\usepackage{bm}
\usepackage[capitalise]{cleveref}
\usepackage[standard]{ntheorem}
\usepackage{tikz}
\usepackage[draft, margin]{fixme}
\fxsetup{theme=color}
\usepackage{graphicx}
\usepackage[algoruled]{algorithm2e}
\usepackage{grffile}
\usepackage[backend=biber,style=numeric,maxbibnames=99]{biblatex}
\usepackage{enumitem}
\usepackage{float}
\usepackage{nicefrac}
\usepackage{csquotes}
\usepackage{booktabs}
\usepackage{textcomp}

\FXRegisterAuthor{pg}{apg}{PG}
\FXRegisterAuthor{mf}{amf}{MF}
\FXRegisterAuthor{fk}{afk}{FK}
\FXRegisterAuthor{aa}{aaa}{AA}

\crefname{assumption}{Assumption}{Assumptions}

\crefname{claim}{Claim}{Claims}
\crefname{equation}{}{}  
\Crefname{equation}{Equation}{Equations}

\DeclareMathOperator\interior{int}

\newcommand{\diag}{\operatorname{diag}}
\newcommand{\vol}{\operatorname{vol}}

\newcommand{\ip}{{\ell}}

\newcommand{\X}{\mathcal{X}}
\newcommand{\Y}{\mathcal{Y}}

\newcommand{\apd}{\mathcal{P}}

\newcommand{\apdmatrices}{\mathcal{A}}
\newcommand{\apdmatrix}{A}
\newcommand{\apdweight}{\gamma}
\newcommand{\apdweights}{\Gamma}
\newcommand{\apdsites}{S}
\newcommand{\apds}{s}

\DeclarePairedDelimiterX{\norm}[1]{\lVert}{\rVert}{#1}
\DeclarePairedDelimiterX{\set}[1]{\lbrace}{\rbrace}{#1}

\newcommand{\R}{\mathbb{R}}
\newcommand{\N}{\mathbb{N}}
\newcommand{\Z}{\mathbb{Z}}

\makeatletter
\newcommand{\leqnomode}{\tagsleft@true\let\veqno\@@leqno}
\newcommand{\reqnomode}{\tagsleft@false\let\veqno\@@eqno}
\makeatother

\pgfplotsset{compat=newest} 
\bibliography{references}

\begin{document}
	\selectlanguage{english}
	\title{Dynamic grain models via fast heuristics for  diagram representations }
	\author{Andreas Alpers, Maximilian Fiedler, \\ Peter Gritzmann, Fabian Klemm}
	
	\publishers{\vspace*{4ex}%
		\normalfont\normalsize%
		\parbox{0.8\linewidth}{%
			\textbf{Abstract.} 
			The present paper introduces a mathematical model for studying dynamic grain growth. In particular, we show how characteristic measurements, grain volumes, centroids, and central second-order moments at discrete moments in time can be turned quickly into a continuous description of the grain growth process in terms of  geometric diagrams (which largely generalize the well-known Voronoi and Laguerre tessellations). We evaluate the computational behavior of our algorithm on real-world data. 
		}	
	}
	
	\maketitle


\section{Introduction}
Grain growth is an important field of study in materials science as the resulting grain structures strongly influence the mechanical and physical properties of metals, ceramics, and other polycrystalline materials~\cite{poulsenbook,graingrowthbook1}. Although a number of different models have been introduced over the years, starting with works of  Smith, von Neumann, and Mullins in the 1950s~\cite{model0,model1,model2}, understanding and controlling grain growth remains challenging, both in theory and practice. For instance, the recent empirical study~\cite{Zhang2018} demonstrates that neither the classical Hillert's model~\cite{model5} nor the MacPherson-Srolovitz model~\cite{model3} capture anisotropic grain growth adequately. Additional challenges are discussed in~\cite{graingrowthchallenges2,modelnew,graingrowthchallenges1}. 

The present paper introduces a mathematical interpolation/extrapolation model for quickly turning information acquired at discrete moments in time into a continuous description of the grain growth process. It employs diagram representations of grain maps (which will be formally introduced in \cref{sec:diagrams}) based on measured estimates of the grain volumes, centroids, and, if available, central second-order moments. Such characteristics can be obtained, for instance, by utilizing tomographic techniques; see~\cite{poulsenbook, intcrystbook}.

The diagram representations considered in this paper are generalizations of  previously used (isotropic) Laguerre and Voronoi tessellations~\cite{Lyckegaard2011, telley96b,Telley1992} and allow for curved boundaries and non-convex grains. While these representations are based on a phenomenological model whose physical foundation is still not fully understood, they have consistently been reported to capture the physical principles governing the forming of polycrystals quite well; see \cite{Alpers2015,Altendorf2014,curvedgrainboundaries,sedivy16, sedivy17, Spettl2016, Teferra2015, teferra18}. Several sophisticated optimization algorithms are available for computing or approximating such diagrams with high accuracy, \cite{Alpers2015,curvedgrainboundaries}, but they require relatively large computational cost. This can be a bottleneck, particularly in dynamic 3D studies. We therefore focus in the present paper on extremely fast optimization-free heuristics for fitting  diagrams to measured data. Methods for the related but generally different task of generating synthetic grain structures that follow a given distribution are discussed in~\cite{barker16,fastLaguerre, guilleminot11, kuehn08,Xu2009}.

First, we use discriminant analysis to interpret a heuristic, subsequently referred to as H1, which was introduced in \cite{Lyckegaard2011} for the isotropic case and extended by \cite{teferra18} to the anisotropic case. Our analysis leads naturally to an even simpler and faster heuristic~H2, proposed  in~\cite{Altendorf2014}, which can now be explained to perform equally well under certain conditions. Also, the comparison of H1 and H2 suggests that diagram parameters may provide additional insight into the stage of the growth process. We apply these heuristics to available grain maps in order to assess their quality.  Then we introduce a dynamic grain growth model to obtain a continuous description of the process based on measurements at only a few moments in time. Finally, we evaluate the computational behavior of our algorithm on real-world data. Our study indicates that the growth process converges towards particularly simple diagrams, which seem to be energetically favorable. 

The paper is organized as follows. \Cref{sec:diagrams}  introduces the relevant diagrams and formalizes the concept of grain scans. Sections \ref{sec:approximation_methods} and \ref{sec:evaluation_H1-H2} address optimization-free heuristics, both, in terms of a theoretical analysis and a practical evaluation on real-world data. 
\cref{sec:dynamic_grain_model} introduces a dynamic model which is capable of extending measurements at a few time steps to a continuous grain growth model very quickly. The results of our empirical study on practical grain data are then given in \cref{sec:evaluation_polycrystals}.  \Cref{sec:conclusion} concludes with some final remarks.

\section{Diagrams and grain scans}\label{sec:diagrams}

Adequate representations of grain maps can be valuable tools for understanding the physical principles of grain growth and the properties of the resulting material. In fact, the diagram representations studied in this paper require only a few parameters per grain and are thus relatively easy to process. As they exhibit geometric and combinatorial features of grain maps which are not readily available otherwise, such representations may contribute to predicting and, potentially, even controlling the forming of new materials.

Some or all parameters which specify a diagram can be estimated from certain tomographic measurements, and optimizing the remaining ones then constitutes an inverse problem; see~\cite{AFGK21b}.
Previous studies~\cite{ Alpers2015, Altendorf2014,curvedgrainboundaries, sedivy16, sedivy17,Spettl2016,Teferra2015, teferra18} indicate that, in particular, grain volumes, centroids, and central second-order moments govern decisive properties of grain maps.  

Let us now formally introduce the appropriate class of diagrams. Let $d\in \N$ denote the \emph{dimension} of space. Here, we are mainly interested in the 3 dimensional case. However, our approach works for general $d$, and, in fact, planar sections or surfaces of polycrystals are also relevant in practice. Further, let $k\in \N$ denote the number of different grains in the underlying grain map. An \emph{anisotropic power diagram} is specified by the following data for each $i\in [k]:=\{1,\dots,k\}$: 
\begin{itemize}
    \item a positive definite symmetric matrix $\apdmatrix_{i}\in \R^{d\times d}$,
     \item a \emph{site} $\apds_i \in \R^d$, and a
      \item \emph{size} $\apdweight_i \in \R$.
\end{itemize}
For simpler reference, we collect these parameters in the two families $\apdmatrices=\{\apdmatrix_{1},\dots,\apdmatrix_{k}\}$ and $\Gamma = \{ \apdweight_1,\dots, \apdweight_k\}$, where repetitions of elements are permitted,  and the set $\apdsites = \set{ \apds_1, \ldots, \apds_k}$. The sites specify the positions of the grains while the matrices in $\apdmatrices$ describe characteristics of their shapes. In fact, each grain is equipped with its own \emph{ellipsoidal norm} $\norm{\quad }_{A_i}$ given through $\apdmatrix_i$ by
\begin{align*}
	\norm{ x }_{A_i} = \sqrt{ x^\top A_i x }, \qquad (x \in \R^d).
\end{align*}
The sizes $\apdweight_i$ are used to control the grain volume. 

Given $\apdmatrices$, $\apdsites$ and $\Gamma$, the \emph{anisotropic power diagram} (in the following often simply referred to as diagram)
\begin{align*}
\apd= \apd(\apdmatrices,\apdsites,\Gamma) =\{P_1,\ldots,P_k\},
\end{align*}
is a decomposition of $\R^d$ into \emph{cells} $P_i$ defined by
\begin{align*}
	P_i = \bigl\{ x \in \R^d:  \norm{x-s_i}_{A_i}^2 + \apdweight_i \leq   \norm{x-s_\ell}_{A_\ell}^2 + \apdweight_\ell, ~~\forall \ell \in [k]\setminus \{i\}\bigr\}.
\end{align*}
In particular, two different cells do not share interior points, i.e., $\interior(P_i)\cap \interior(P_\ell)=\emptyset$ for~$i\ne \ell$.
See \cite{Brieden2017} and \cite{Gritzmann2017} for examples and further properties of different classes of diagrams. 

Diagrams can be viewed as  continuous (or resolution-independent) representations of  polycrystalline samples. Conventionally, the samples are represented as voxel-based digital images at some resolution, called  \emph{grain maps}. More generally, given a set of points  $X=\{x_1,\dots,x_n\}\subseteq \mathbb{R}^d$ in the sample and labels, i.e., $y_1,\dots,y_n\in[k]$, with $k\in \N$ denoting the number of different grains in the sample, we will call the labeled data set $\bigl\{(x_j,y_j): j\in [n]\bigr\}$ a \emph{grain scan} or \emph{polycrystal scan}. In the following, we refer to it as $(X,Y)$ with the understanding, that the indices of the labels in the family $Y= \{ y_1,\dots , y_n \}$ correspond to the indices of the points in~$X$.  A grain scan therefore provides a partitioning of $X$ into (the discretizations of) the \emph{grains} $G_i=\bigl\{x_j: (x_j,i)\in (X,Y)\bigr\}$. We can then measure the fit of the diagram in terms of the symmetric difference of $G_i$ and $X \cap P_i$ for all $i \in [k]$. 
(Points on the boundary of two or more cells are assigned according to some preference criterion or, conservatively, can be viewed as classification error, but are essentially irrelevant in our context and hence generally discarded.) 

Usual grain maps are special grain scans where the sample is assumed to be in the range $[0,1]^d$ and points of $X$ are identified with the points of $[0,1]^d$ in the grid $\rho \Z^d$ for some \emph{resolution}  $\rho \in (0,1)$. Note that, with $\kappa_i=|G_i|$, the term $\rho^d \kappa_i$ can be regarded as an approximation of the volume $\nu_i$ of the $i$th grain.

If a grain scan is available the computation of a best fitting diagram requires the optimization of all three characteristics $\apdmatrices$, $\apdsites$ and $\apdweights$. While this can indeed be done, see \cite{AFGK21b}, it is, however, too time-consuming, given current optimization technology, for dynamic 3D applications. Hence, we will in the following assume that $\apdmatrices$ and $\apdsites$ are available through measurements and  $\apdweights$ will be determined by a heuristic which may, additionally, use grain volume information, i.e., (approximations of) $\nu_1,\ldots,\nu_k$. As such heuristics abstain from any optimization, they are computationally fast but potentially result in a higher misclassification error. 

\section{Optimization-free heuristics}
\label{sec:approximation_methods}

This section considers two heuristics that derive all characteristic parameters $\apdmatrices,\apdsites,\apdweights$ for an anisotropic power diagram directly from given measurements. These heuristics avoid optimization routines altogether, are fast and, as we will see, already provide a reasonable classification accuracy. More specifically, we provide a new interpretation of the heuristics introduced in \cite{Lyckegaard2011,teferra18} and \cite{Altendorf2014} within the framework of discriminant analysis.

\subsection*{Choices for $\apdmatrices$ and $\apdsites$}\label{ssec:matrices-sites}
Discriminant analysis is a popular classification technique in statistics and provides a different perspective on the a priori choices for the characteristic parameters $(\apdmatrices,\apdsites,\apdweights$); see e.g.~\cite{learningbook}.  Within its realm, the grain scan is regarded as a sample of the multivariate random variable $(\X,\Y)$ where $\X$ specifies a point in $[0,1]^d$ and $\Y$ gives the index $i\in [k]$ of the corresponding grain. Discriminant analysis estimates the distribution of the multivariate random variable from samples and assigns a point to the class with the highest probability $p$. Hence, the decision boundary between two classes $i,\ip$ consists of those points $x$ that have an equal conditional probability of lying in either class, i.e.,
\begin{equation*}
	\bigl\{x \in \R^d : p(\Y = i|\X=x) = p(\Y = \ip|\X=x)\bigr\}.
\end{equation*}
For its estimation, discriminant analysis applies the maximum likelihood principle under the assumption that the members from each class are normally distributed. As each normal distribution $\mathcal{N}(\mu_i,\Sigma_i)$ is fully determined by its $d$-dimensional mean vector $\mu_i$ and its $(d\times d)$-covariance matrix $\Sigma_i$, it is these characteristics that need to be estimated. It is well known that the maximum likelihood principle results in the choices of the sites~$\apdsites$ as the centroids and the matrices $\apdmatrices$ as the inverse of the covariance matrices of the sample, i.e., more precisely,
\begin{align*}
   \mu_i &= c_i = \frac{1}{|G_i|} \sum_{x_j\in G_i} x_j,\\
   A_i & = \bigl(\Sigma(G_i)\bigr)^{-1} = \left(\frac{1}{|G_i|} (x_j-c_i:x_j\in G_i) (x_j-c_i:x_j\in G_i)^T\right)^{-1}
\end{align*}
for $i\in [k]$.
Note that the latter requires that the matrices $\Sigma(G_i)$ have full rank $d$. This can be assumed here as, in the underlying model, grains that are degenerate to volume~$0$ do not matter. The above maximum likelihood setting is precisely the choice for $\apdmatrices$ and $\apdsites$ suggested in \cite{Alpers2015}, and subsequently also used in \cite{teferra18}. Also, estimates for these parameters are available through tomographic measurement.

Let us point out that, in practice, discriminant analysis is often applied even when the underlying assumption of a normal distribution is only met approximately (as is the case here). This is justified as the procedure is generally reported to be quite robust; see, e.g., \cite{Lachenbruch1979}. 

\subsection*{Choices for $\apdweights$}\label{ssec:gamma}
We will now turn to the third set of characteristics, i.e., the sizes $\apdweights$. We show first that the choice of $\apdweights$ in heuristic H1 can also be interpreted within the paradigm of discriminant analysis with a specific prior. Building on this interpretation, a different natural choice will lead to H2.

To begin with, let us briefly recall the volume argument of \cite{Lyckegaard2011} and (in its anisotropic generality) \cite{teferra18} that leads to the specific choice for $\apdweights$ in heuristic H1. 

Let $A\in \R^{d\times d}$ be positive definite and symmetric. As $A$ can be diagonalized there exist an orthogonal matrix $U\in \R^{d\times d}$ and a diagonal matrix $D=\diag(\lambda_1,\ldots,\lambda_d)\in \R^{d\times d}$ with $\lambda_1,\ldots,\lambda_d>0$ such that $A=UDU^T$. Then, using the abbreviation $D^{\nicefrac{1}{2}}=\diag(\sqrt{\lambda_1},\ldots,\sqrt{\lambda_d})$, we have for the unit ball $\mathbb{B}^d_{A}$ and $\rho\in (0,\infty)$
\begin{align*}
  \rho \mathbb{B}^d_{A} & = \bigl\{x\in \R^d: \norm{x}_{A} \le \rho \bigr\}=
   \bigl\{x\in \R^d: x^TAx \le \rho^2 \bigr\} \\
   &= \bigl\{x\in \R^d: x^TUDU^Tx \le \rho^2 \bigr\}
    =\rho UD^{-\nicefrac12} \mathbb{B}^d_{2}.
\end{align*}
Hence, the volume $\vol\bigl(\rho \mathbb{B}^d_{A}\bigr)$ of this ellipsoid is given by
\begin{equation*}
  \vol\bigl(\rho \mathbb{B}^d_{A}\bigr) = \rho^d \vol\bigl(UD^{-\nicefrac12} \mathbb{B}^d_{2}\bigr) = \rho^d \det(A)^{-\nicefrac12} \vol\bigl(\mathbb{B}^d_{2}\bigr).
\end{equation*}
Regarding the ellipsoid $c_i+\rho \mathbb{B}^d_{A_i}$ as an approximation of the $i$th grain, it is now natural to choose $\gamma_i$ according to the volume condition
\begin{equation*}
    \vol \left(\bigl\{x\in \R^d: \norm{x}_{A_i}^2 + \gamma_i \le 0 \bigr\}\right) = \nu_i
\end{equation*}
involving the available approximation $\nu_i$ for the volume of the $i$th grain. A simple computation yields
\begin{equation*}
    \gamma_i = -\rho^2 = -\left(\frac{\nu_i}{\vol\bigl(\mathbb{B}^d_{A_i}\bigr)}\right)^{\nicefrac2d}= -\left(\frac{\nu_i \sqrt{\det(A_i)}}{\vol\bigl(\mathbb{B}^d_{2}\bigr)}\right)^{\nicefrac2d}
\end{equation*}
For $d=3$, and with the specific setting of $A_i=\bigl(\Sigma(G_i)\bigr)^{-1}$, this yields the choice 
\begin{equation*}
    \gamma_i = 
    - \left(\frac{3\nu_i}{4\pi \sqrt{\det(\Sigma(G_i))}}\right)^{\nicefrac23}
\end{equation*}
from~\cite{Lyckegaard2011, teferra18}.
In order to interpret this choice via discriminant analysis, we determine which \emph{probability distribution}, called  \emph{prior}, will lead to the same decision boundaries between the classes. 
So, let 
\begin{equation*}
\pi_1,\dots, \pi_k\in [0,\infty), \qquad N=\sum_{i=1}^k \pi_i \ne 0.
\end{equation*}
Then $(\pi_1,\dots, \pi_k)$ specifies the a priori \enquote{belief} about the \emph{prior probability} $\pi_i/N$ that a point (regardless of its position) belongs to cluster~$i$, i.e., 
\begin{equation*}
    \frac{\pi_i}{N}=p(\Y=i), \qquad (i\in [k]).
\end{equation*}
The computation of the decision boundary is then facilitated via \emph{Bayes' Theorem}. As each class is assumed to be normally distributed, Bayes' Theorem involves the conditional density functions
\begin{equation*}
    f_{\X|\Y=i}(x)= \frac{\det(A_i)^{\nicefrac{1}{2}}}{(2\pi)^{\nicefrac{d}{2}}}  \cdot \exp\left(-\frac{1}{2}(x-c_i)^\top A_i (x-c_i)\right)
\end{equation*}
of the $k$ estimated normal distributions $\mathcal{N}(\mu_i,\Sigma_i),$ i.e., with
\begin{equation*}
    \mu_i=c_i, \quad \Sigma_i=\Sigma(G_i)=A_i^{-1}.
\end{equation*}
In this situation, Bayes' Theorem states that
\begin{equation*} 
	p(\Y=i|\X=x) = \frac{f_{\X|\Y=i}(x)\pi_i}{\sum_{\ell=1}^k f_{\X|\Y=\ell}(x) \pi_\ell}.
\end{equation*}
Therefore, the maximum likelihood decision boundaries between two classes $i,\ip$ can be determined by solving the equation 
\begin{equation*}
f_{\X|\Y=i}(x)\cdot \pi_i=f_{\X|\Y=\ip}(x)\cdot \pi_{\ip}.
\end{equation*}
Taking the logarithm, using that $(x-c_i)^\top A_i (x-c_i)=\norm{x-c_i}_{A_i}^2$, and simplifying, this condition turns into the equation
	\begin{equation*}
\norm{x-c_i}_{A_i}^2 -\log\bigl(\det(A_i)\bigr) - 2\log(\pi_i)
		= \norm{x-c_\ip}_{A_\ip}^2 - \log\bigl(\det(A_\ip)\bigr) - 2\log(\pi_\ip).
	\end{equation*} 
Hence, these decision boundaries coincide with that of the diagram with the parameters $(\apdmatrices,\apdsites,\apdweights)$ if and only if
\begin{equation*}
	 \apdweight_i =\apdweight_i^{(1)}= -\log\bigl(\det(A_i)\bigr) - 2\log(\pi_i)+ \gamma, \qquad (i\in[k]),
\end{equation*}
where  $\gamma\in\mathbb{R}$ is an arbitrary constant. (Recall that the diagrams are invariant under such common additions to the sizes.) For the above choice of the sizes in H1, we hence obtain 
\begin{equation*}
	\pi_i = \pi_i^{(1)}= \frac{e^{\nicefrac{\gamma}{2}}}{\sqrt{\det(A_i)}}  \cdot \exp\left(\frac{1}{2}\left(\frac{\nu_i \sqrt{\det(A_i)}}{\vol(\mathbb{B}^d_{2}) }\right)^{\nicefrac{2}{d}}\right), \qquad (i\in[k]).
\end{equation*}
Since
\begin{equation*}
    \frac{1}{\sqrt{\det(A_i)}} = \sqrt{\det(\Sigma(G_i))}= \frac{1}{\vol(\mathbb{B}^d_{2})}\cdot \vol(\mathbb{B}^d_{A_i}),
\end{equation*} 
the first factor is proportional to the volume of the ellipsoid $\mathbb{B}^d_{A_i}$. The second factor is a \enquote{correction term} which increases exponentially if the measured grain volume $\rho^d\kappa_i$ exceeds $\vol(\mathbb{B}^d_{A_i})$. The first term seems natural while the second is quite unusual. 

Within the realm of discriminant analysis it appears reasonable to simply set $\apdweight_i  = 0$ for all clusters $i$. Then we obtain  the priors
\begin{equation*}
	\pi_i = \pi_i^{(2)}=  \frac{e^{\nicefrac{\gamma}{2}}}{\sqrt{\det(A_i)}},\qquad (i\in[k]),
\end{equation*}
which are proportional to the volumes of the ellipsoids corresponding to the covariance matrices of the grains. Thus, this simpler prior distribution encodes the assumption that the ellipsoid volumes already represent the grain volumes well. This justifies the choice \begin{equation*}
\apdweight_1^{(2)}=\ldots =\apdweight_k^{(2)}=0,
\end{equation*} 
proposed in~\cite{Altendorf2014}. We will refer to the respective heuristic as H2. 

In the next section we compare the behavior of H1 and H2 on available real-world grain scans.

\section{Comparison of H1 and H2}\label{sec:evaluation_H1-H2}

We compare the two heuristics H1 and H2 on a real-world 3D grain scan and indicate how their difference can be interpreted. This may be of particular interest for understanding the state of a grain growth process and will be taken up in \cref{sec:evaluation_polycrystals}.

\subsection*{Experimental evaluation}
We compare the methods with a data set taken from \cite{Lyckegaard2011} that has been obtained by a synchrotron micro-tomograph experiment conducted on a metastable beta titanium alloy (Ti $\beta$21S). The material was scanned with the volume of 240\textmu m $\times$ 240\textmu m $\times$ 420\textmu m at a resolution corresponding to a voxel size of 0.7 \textmu m. This scan resulted in a 3D voxel image of size 339 $\times$ 339 $\times$ 599 composed of a total of 591 grains. All our computations are carried out in 3D, i.e., $d=3$. 

In the following, we assess the quality of fit by means of different measures. Let $\apd=\{P_1,\ldots,P_k\}$ be a diagram (obtained by H1, H2, or, for that matter, any other algorithm). We set  
\begin{equation*}
  \hat{C}_i= \bigl\{x_j\in X: x_j \in \interior(P_i)\bigr\} \quad \bigl(i\in [k]\bigl), \qquad \hat{C}=(\hat{C}_1,\ldots,\hat{C}_k).
\end{equation*}
Note that $\hat{C}$ is not a clustering of $X$ but only of $X\cap \bigcup_{i=1}^k \interior(P_i)$; this is indicated by writing $\hat{C}$ rather than $C$. 

A natural measure how well $\apd$ captures the grain scan $G=(G_1,\ldots,G_k)$ is the \emph{relative fit} or \emph{accuracy} $\Phi_G$ of $\hat{C}$ defined by
\begin{equation*}
  \Phi_G(\hat{C})= \frac{1}{n} \left| \bigcup_{i=1}^k \bigl(G_i\cap \hat{C}_i\bigr) \right|.
\end{equation*} 
Let us point out that, as they are not assigned within $\hat{C}$, all points on the cell boundaries count as misclassified. 

In \cref{tb:comparison_approximation_methods} we will also report on the behavior with respect to the \emph{relative cluster weight error} 
\begin{equation*}
	\Psi_G(\hat{C}) = \frac{1}{n}\sum_{i=1}^k\bigl|\kappa_i - |\hat{C}_i)|\bigr|,
\end{equation*}
and the relative deviation of the centroids $c(G_i)$ and $c(\hat{C}_i)$, and the covariance matrices $A_i^{-1}= \Sigma_i=\text{Cov}(G_i)$ and $\text{Cov}(\hat{C}_i)$, respectively. Formally, the \emph{relative centroid error} and the \emph{relative covariance error} are defined as
\begin{equation*}
\frac{1}{n} \sum_{i=1}^k \kappa_i\norm{c(G_i) - c(\hat{C}_i)}_{2} \quad \text{and} \quad
 \frac{1}{n}\sum_{i=1}^k \kappa_i\norm{\Sigma_i - \text{Cov}(\hat{C}_i)}_2.
\end{equation*}
For the latter, we used the spectral norm. This matrix norm measures the largest singular value of the covariance differences and, thus, intuitively captures the difference in the largest singular vectors of the covariance matrices. Note that, as defined, the latter two measures depend on the original physical dimensions of the sample. We refrain from any additional normalization as we do not conduct inter-sample studies here.

We will also address combinatorial features. More precisely, we consider the percentage of grains with a correct neighborhood, i.e., with all grain neighbors in the ground truth being also neighbors in the representation. Here, a grain is a neighbor of another grain if a voxel of the former grain is adjacent (face-connected, edge-connected or vertex-connected) to a voxel of the later grain. We also provide this percentage when one and, respectively, two errors in the neighborhood are allowed, i.e., when there is either one additional or one missing voxel. 

The results of our evaluation are shown in \cref{tb:comparison_approximation_methods}.

\begin{table}[H]
	\centering
	\begin{tabular}{lrr}
		\hline
		\textbf{Performance Characteristics}      	& \textbf{H1}    & \textbf{H2}  \\ \hline
	  Accuracy (in \%)         & 92.98652         & 92.98658         \\
Relative Cluster Weight Error  & 0.02869       & 0.02869    \\
Relative Centroid Error (in $\mu$m)      & 0.64346          & 0.64347        \\
Relative Covariance Error (in $\mu$m$^2$)    & 15.25372         & 15.25376 \\
Correct Neighborhoods (in \%) 	& 42.30118      & 42.30118    \\
Correct Neighborhoods up to 1 Error (in \%)  & 80.54146          & 80.54146    \\
Correct Neighborhoods up to 2 Errors (in \%)  & 94.92386         & 94.92386    \\
 Running Time (in s)       	& 1.49641           & 0.00000 \\ \hline
		\end{tabular}
		\caption{Comparison of H1 and H2.}
		\label{tb:comparison_approximation_methods}
		\end{table}

Let us point out that the displayed running times involve the heuristics but, for a fairer comparison, do not include the computation of the performance criteria. Hence, the entry $0$ for the running time of H2 refers to the fact that all parameters defining $\apd$ are given upfront. 
Also, while we exercised appropriate care for the implementations, we do not claim that our code is fully optimized. 

In terms of fit, we observe that both heuristics perform extremely well, in fact, almost identical, on the given real-world data set. We will now offer an explanation for the latter behavior.

\subsection*{Similarity and difference}\label{sec:difference}

While both heuristics, H1 and H2, employ the same matrices $\apdmatrices$ and sites $\apdsites$, they generally differ in the sizes $\apdweights$ of the resulting diagrams. Hence, the nearly identical performance depicted in \cref{tb:comparison_approximation_methods} for all applied quality criteria may come as a surprise. 
As observed before, the priors  underlying H1 and H2 denoted $\pi_i^{(1)}$ and~$\pi_i^{(2)},$ as before, differ by a factor that depends exponentially on the deviation of $\vol(\mathbb{B}^d_{A_i})$ from the measured grain volume $\nu_i$. 

Now, let $\alpha_i$ be a positive real number such that
\begin{equation*}
  \nu_i=\alpha_i \cdot \vol(\mathbb{B}^d_{A_i}) = \alpha_i\cdot \frac{\vol(\mathbb{B}^d_{2})}{\sqrt{\det(A_i)}}.
\end{equation*} 
Then, for each $i\in [k]$,
\begin{align*}
    \pi_i^{(1)} & = \frac{e^{\nicefrac{\gamma}{2}}}{\sqrt{\det(A_i)}}  \cdot \exp\left(\frac{1}{2}\left(\frac{\nu_i \sqrt{\det(A_i)}}{\vol(\mathbb{B}^d_{2}) }\right)^{\nicefrac{2}{d}}\right) = \frac{e^{\nicefrac{\gamma}{2}}}{\sqrt{\det(A_i)}}  \cdot \exp\left(\frac{1}{2} \alpha_i^{\nicefrac{2}{d}}\right) \\
    & = \pi_i^{(2)} \cdot \exp\left(\frac{1}{2} \alpha_i^{\nicefrac{2}{d}}\right),
\end{align*}
and, expressed in terms of the sizes $\gamma_i^{(1)}$ used in H1, we obtain
\begin{equation*}
    \apdweight_i^{(1)}=-\left(\frac{\nu_i}{\vol\bigl(\mathbb{B}^d_{A_i}\bigr)}\right)^{\nicefrac2d}=-\alpha_i^{\nicefrac{2}{d}}.
\end{equation*}

Now, suppose that all factors $\alpha_i$ are identical, i.e., $\nu_i=\alpha \cdot \vol(\mathbb{B}^d_{A_i})$ for some positive real number $\alpha$ independent of~$i.$ Then, of course,
\begin{equation*}
 \pi_i^{(1)}= \pi_i^{(2)} \cdot \exp\left(\frac{1}{2} \alpha^{\nicefrac{2}{d}}\right), \quad
 \apdweight_i^{(1)}=-\alpha^{\nicefrac{2}{d}}, \qquad\qquad (i\in [k]).
\end{equation*}
Hence, the (normalized) priors for H1 and H2 coincide, which implies that the maximum likelihood decision boundaries for the two heuristics are the same. Further, since diagrams are invariant under the addition of a constant to each size, the diagrams produced by H1 and H2 are identical.

We suspect that grain structures for which the volumes of the unit balls $\mathbb{B}^d_{A_i}$ with respect to $A_i=\bigl(\Sigma(G_i)\bigr)^{-1}$ are nearly proportional to the grain volumes $\nu_i$ might turn out to be energetically favorable. According to the above reasoning, the former may be quantified by the variance of the powers $\Gamma^{\nicefrac{d}{2}}=(\gamma_1^{\nicefrac{d}{2}},\ldots,\gamma_k^{\nicefrac{d}{2}})$ computed in H1, i.e.,
\begin{equation*}
 \operatorname{var}(\Gamma^{\nicefrac{d}{2}})= E\bigl(\Gamma^{\nicefrac{d}{2}}-E(\Gamma^{\nicefrac{d}{2}})\bigr)=\frac{1}{k} \sum_{i=1}^k \left( \gamma_i^{\nicefrac{d}{2}}-\frac{1}{k} \sum_{i=1}^k  \gamma_i^{\nicefrac{d}{2}}\right)^2.
\end{equation*}
As an example, we will in \cref{sec:evaluation_polycrystals} compute the evolution of $\operatorname{var}(\Gamma^{\nicefrac{3}{2}})$ at different moments in time for a real-world 3D grain growth process.
In terms of H1 and H2, we might expect that if the two heuristics perform differently in practice, the current grain structure will not be energetically minimal, and further growth will result in an improved structure.

We close this section with a \enquote{constructed} theoretical example which sheds some additional light on the behavior of the two heuristics. Since the general principle of the construction is the same in higher dimension and can, hence, already be illustrated in 2D we restrict the following description to the case $d=2$.  

\begin{figure}[ht]
	\centering
		\includegraphics[width=\textwidth]{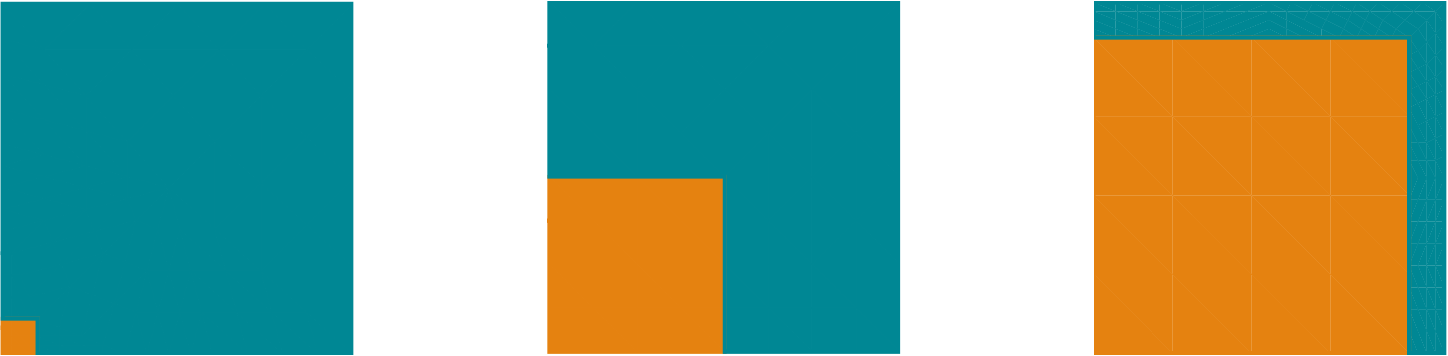}\vspace*{2ex}
		\includegraphics[width=\textwidth]{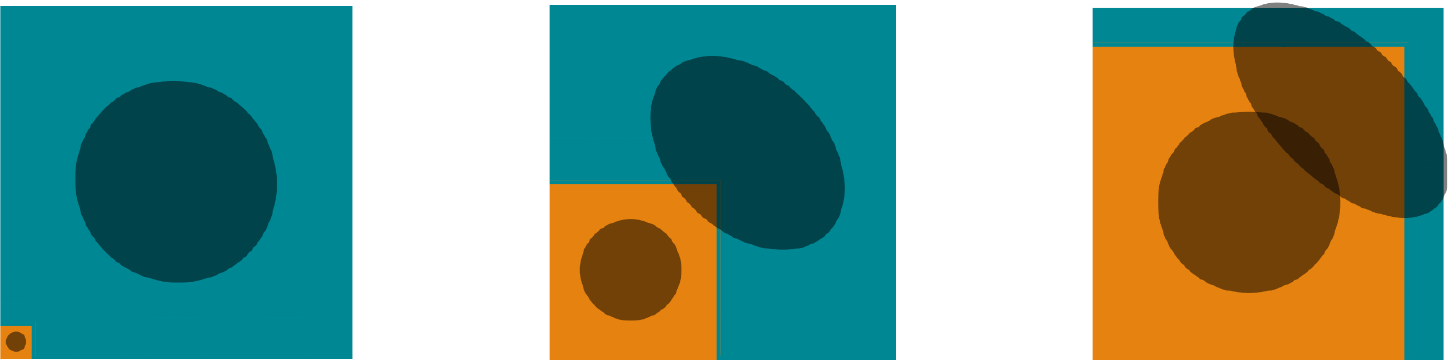}
		\caption{A family of grain scans involving a parameter $\beta$ to illustrate differences between H1 and H2. Top, from left to right: Grain scans for $\beta=0.1$, $\beta=0.5$, and $\beta=0.9$. Grain $G_1$ (orange) fills the square $[0,\beta]\times[0,\beta]$, $G_2$ (blue) fills the rest of the unit square. The bottom row shows the computed ellipses $c_i+\mathbb{B}^2_{A_i}$ with  $A_i=\bigl(\Sigma(G_i)\bigr)^{-1}$ for $i=1,2$.}
		\label{fig:counterexample_series}
\end{figure}

As depicted in the top row of \cref{fig:counterexample_series}, the image consists of two grains $G_1$ and $G_2$; one (orange) is the square $[0,\beta]\times [0,\beta]$ for different values of $\beta \in (0,1)$, the other (blue) is its complement in $[0,1]\times [0,1].$  Of course, one would not expect such samples to originate from real-world grain scans. The bottom row of~\cref{fig:counterexample_series} shows the corresponding ellipses $c_i+\mathbb{B}^2_{A_i}$ with respect to $A_i=\bigl(\Sigma(G_i)\bigr)^{-1}$. 

\begin{figure}[ht]
	\centering
		\includegraphics[width=\textwidth]{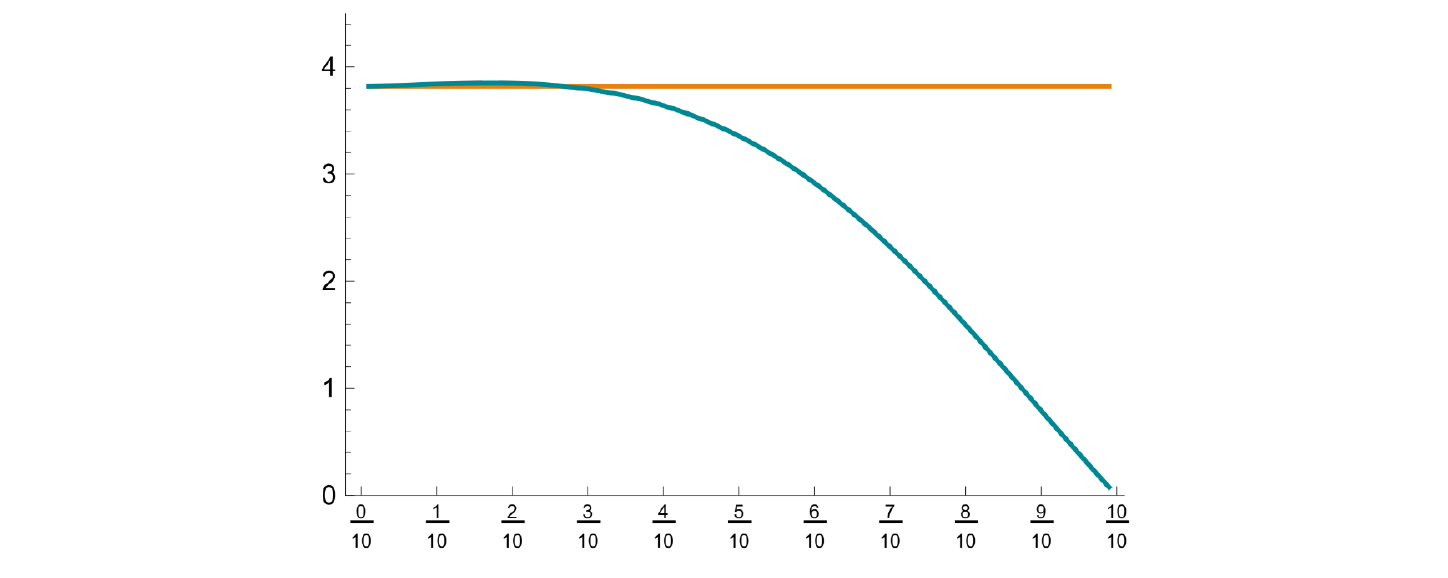}
		\caption{Ratios $\nu_i/\vol\bigl(\mathbb{B}^2_{A_i}\bigr)$ of the areas of the grain and the corresponding ellipsoid for $i=1$ (orange) and $i=2$ (blue) as a function of~$\beta$.}
		\label{fig:counterexample_series-3}
\end{figure}

As argued before, the variance of the ratios $\nu_i/\vol\bigl(\mathbb{B}^2_{A_1}\bigr)$ is a relevant parameter. \Cref{fig:counterexample_series-3} therefore provides  plots of 
\begin{equation*}
    \frac{\beta^2}{\vol\bigl(\mathbb{B}^2_{A_1}\bigr)} \quad \text{and} \quad
    \frac{1-\beta^2}{\vol\bigl(\mathbb{B}^2_{A_2}\bigr)}
\end{equation*} 
as a function of $\beta$.

\begin{figure}[ht]
	\centering
		\includegraphics[width=\textwidth]{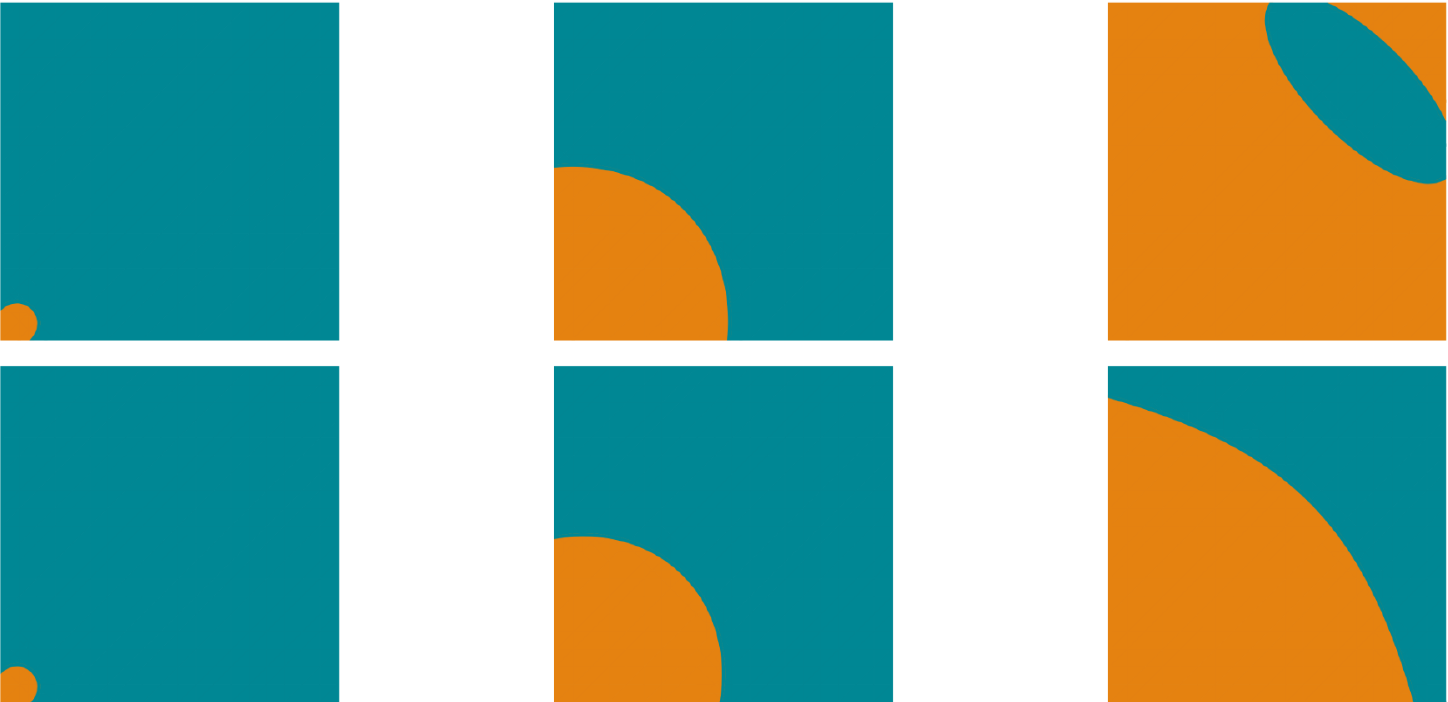}
		\caption{Diagrams for the family of grain scans from~\cref{fig:counterexample_series} obtained via H1 (top row) and H2 (bottom row).}
		\label{fig:counterexample_series-4}
\end{figure}

Note that these ratios differ considerably for larger values of $\beta$. Hence, we would expect that the heuristics H1 and H2 behave quite differently in these cases. This is confirmed by the resulting diagrams depicted in \cref{fig:counterexample_series-4}.

The behavior for $\beta=0.9$ suggests that H1 tries to preserve the areas of the grains, while H2 seems to favor their shapes. If the relevant volume ratios are close, as is the case for small $\beta$, both heuristics achieve both aims and perform similarly.

\begin{figure}[ht]
	\centering
		\includegraphics[width=1\textwidth]{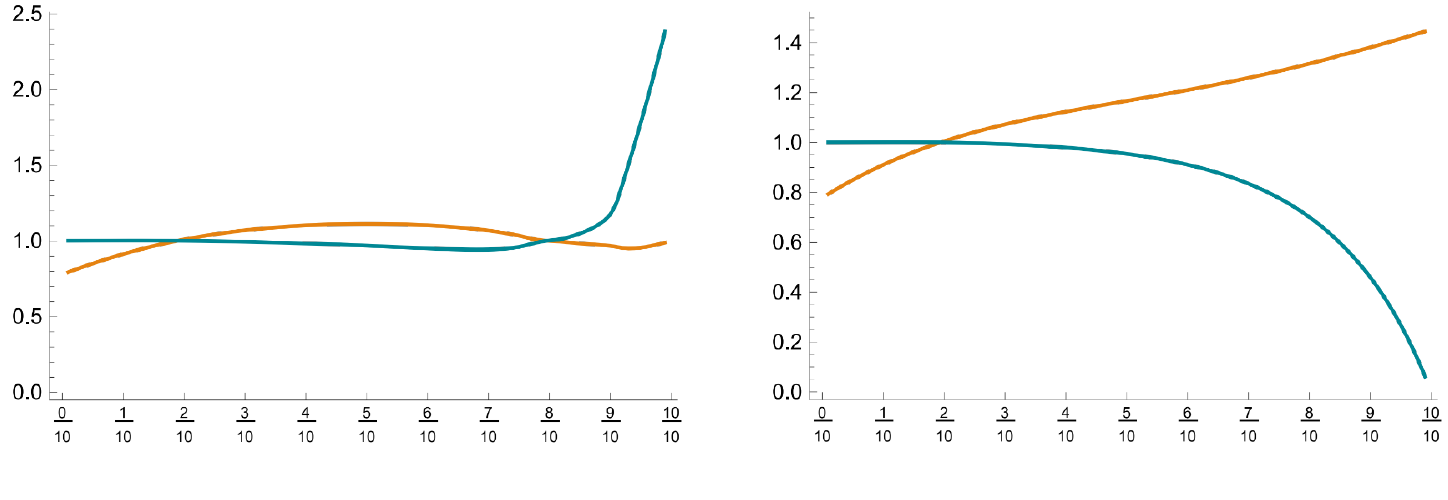}
		\caption{
Ratio of the area $\nu_i$ of the original grain  and the area of the corresponding diagram cell obtained by H1 (left) and H2 (right) for $i=1$ (orange) and $i=2$ (blue) as a function of~$\beta$ . }
		\label{fig:counterexample_series-5}
\end{figure}

\Cref{fig:counterexample_series-5} shows the ratios of the grain volume $\nu_i$  and the areas of the diagram cells under the two models H1 and H2, respectively.

\section{Dynamic grain growth}
\label{sec:dynamic_grain_model}

We now describe a simple interpolation/extrapolation model that is capable of quickly turning information acquired at a few moments in time into a continuous description of the grain growth process. 
Then we report on the results of an empirical evaluation on a real-world grain scan time series. Finally, we study the correlation between the diagram size parameters $\apdweights$ and the disappearance of grains in the process.

\subsection*{The model}\label{sec:model}
Since well-chosen diagrams have been demonstrated to yield high-quality representations of grain structures of many types of  polycrystalline materials, it is natural to exploit their rather low-dimensional parameter space for deriving continuous models from data measured at discrete times. Recall that a diagram is fully characterized by the triple~$(\apdmatrices,\apdsites,\Gamma)$. In order to simulate the dynamic evolution of grain structures over time, we can fit a polynomial (or any other desired type of function) for each diagram parameter extracted from a discrete-time series polycrystal data set. These functions then represent the continuous change in diagram parameters during the growth process. Understanding the evolution of the diagram parameters over time may help predict changes in the material during an annealing process and may provide further insights into the material characteristics. 

Suppose, that 3D-grain scans (or at least appropriate characteristic measurements) are available at the discrete times $t_0,t_1,\ldots,t_m$. Then we compute, for each $t=t_\ell$, the parameters of an anisotropic power diagram $\apd_t=\apd(\apdmatrices_t,\apdsites_t,\Gamma_t)$ which represent the grain scans. Subsequently, we select a class $\mathcal{F}$ of parameterized functions and determine one that best fits the data $\apdmatrices_t,\apdsites_t,\Gamma_t$. In our experimental setting we choose polynomials of a small degree to avoid overfitting, and we measure the fit in terms of least-squares deviation. 

In 3D, the interpolation/extrapolation involves ten functions of $\mathcal{F}$ for each grain, three for the coordinates of the diagram sites, three for the rotation part of the grains' covariance matrix, three for the length of the principal axis of the covariance matrix, and one for the size parameter. Of course, in order to set up the computation of the functions, we need to \enquote{track} each grain, i.e., we need to identify which grain corresponds to which at the different moments $t$ in time. 

During the grain growth process, grains may disappear. We assume, however, that no new grains emerge. In the experimental series that will be described in more detail below, about $40\%$ of all initially present grains disappeared throughout the process. Hence we distinguish between such \emph{non persistent} grains and \emph{persistent} grains that exist throughout the observed time interval. 
Of, course, for non persistent grains, the functions are fitted only as long as the grain \enquote{definitely exists}. More specifically, if a previously existing grain disappears from the measurement, we choose as last diagram parameters the covariance matrix from the previous time step and compute a size parameter such that its grain volume is essentially zero. If, on the other hand, for some moment in time our prediction generates a covariance matrix and a size parameter which result in a grain volume close to zero, we exclude the grain from further consideration.

The diagrams $\apd_t$ can, in principle, be computed by any suitable method. In addition to the obvious criteria that any employed algorithm should be computationally fast and produce diagrams of appropriate accuracy, one has to pay attention to its behavior concerning the invariances within the parameter space to avoid corruption of the interpolation/extrapolation. As we are using the heuristic H1 described in \cref{sec:approximation_methods}, all parameters are, however, uniquely determined by the measurements, and we can hence, in the following, ignore this issue. 

\subsection*{The data}\label{sec:data}
The data used for our experimental study is taken from \cite{Zhang2018}. It comprises 15 3D scans of a 99.9\% pure polycrystalline iron sample taken at different moments in time of an annealing process. More precisely, the material was first cold-rolled and annealed for 30 minutes at 700\textdegree C to fully recrystallize before it was first scanned (time $t_0)$. Subsequently, the sample was annealed again at 800\textdegree C for 10 minutes to simulate grain growth, cooled to stabilize, and then scanned again (time $t_1)$. This was repeated another 13 times with annealing periods of 5 minutes (times $t_2,\dots,t_{14}$). The effect of the process is already visible by the number of grains in the sample. While the first scan exhibits~1327 grains, the last one contains only 776, and the number of interior grains strictly decreases from~387 to 189; see \cite[Table 1]{Zhang2018} for more details. The identification of grains over the~15 moments in time is already provided by \cite{Zhang2018}.

For this data set, we computed 15 anisotropic power diagrams. We used H1 as we were also interested in featuring the temporal development of the corresponding sizes in view of the relation of H1 and H2 elaborated in \cref{sec:approximation_methods}.

\subsection*{Evaluation}\label{sec:evaluation_polycrystals}

We apply a cross-validation to assess the quality of the model. To compute the accuracy at time $t_\ell$, we exclude the data for $t_\ell$ and fit the functions on the data for the remaining times $\{t_0,\ldots,t_{14}\}\setminus \{t_\ell\}$. Then we evaluate the obtained interpolation/extrapolation functions at time $t_\ell$ to compute diagram parameters $\hat \apdmatrices_\ell,\hat \apdsites_\ell,\hat \Gamma_\ell$. Finally, we determine the accuracy $\Phi$ of the resulting diagram $\hat \apd_\ell$ of representing the grain scan at $t_\ell$.
 
\begin{figure}[ht]
    \centering
    \includegraphics[width=1\textwidth]{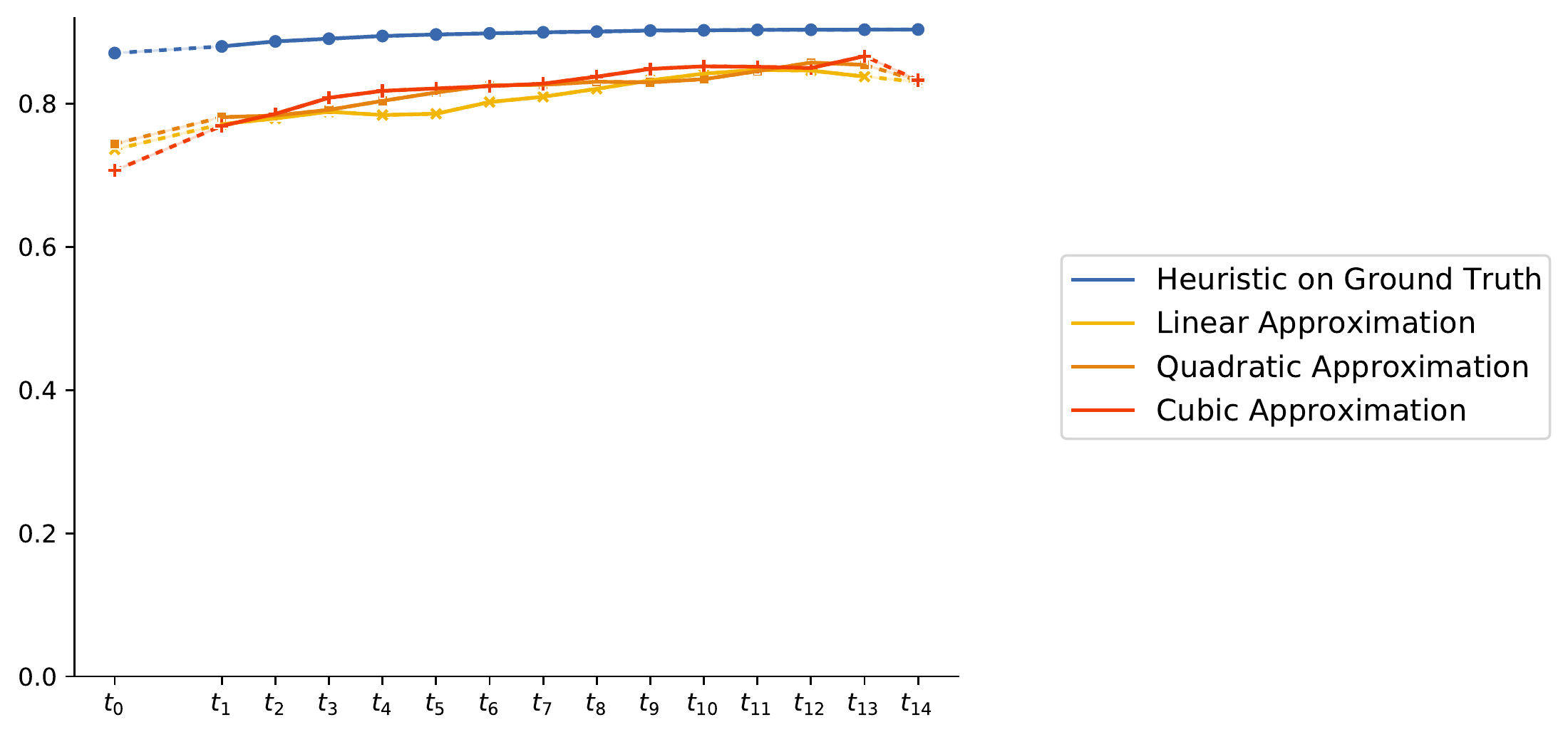}
    \caption{Accuracy (within the cross-validation paradigm) of the  interpolation/extrapolation model plotted for each time step using polynomials of degree $1$, $2,$ and $3,$ respectively for the fit.}
    \label{fig:accuracy_simulation}
\end{figure}

The results for polynomials of degree $1$, $2$, and $3$ are depicted in \cref{fig:accuracy_simulation} in yellow, orange, and red, respectively. Let us point out that the piecewise linear curves which connect the 15 data points for each color are only drawn for a better perception of the trends over time. 

We observe that the degree of the employed polynomials does not seem to have a strong impact on the fit. It would be interesting to link this observation to the \enquote{uniform setup} in terms of the actions over time of the underlying grain growth process. 

As an additional reference the blue markers give an upper \enquote{orientation} for the accuracy that one may expect from the model. In fact, the blue markers depict the accuracy of H1 at each of the 15 time steps.
The curves are dashed at both ends to signify that the model changes from interpolation to extrapolation at these extremes. Also, note that the earlier annealing times differ from the latter ones.

\Cref{fig:accuracy_simulation} shows that the accuracy of our models is generally high. With quadratic polynomials, we obtain an initial accuracy of~74\%, which increases over time to approximately~86\% for later time steps. Note that this is not much below the accuracy obtained by H1 on the ground truth. The general trend that accuracy increases with time (and hence prediction becomes easier) for this data set is in perfect agreement of the findings in ~\cite{Zhang2018, Zhang2020}, which show that over time the system approaches the self-similar regime.   

\begin{figure}[h!]
    \centering
    \includegraphics[width=1\textwidth]{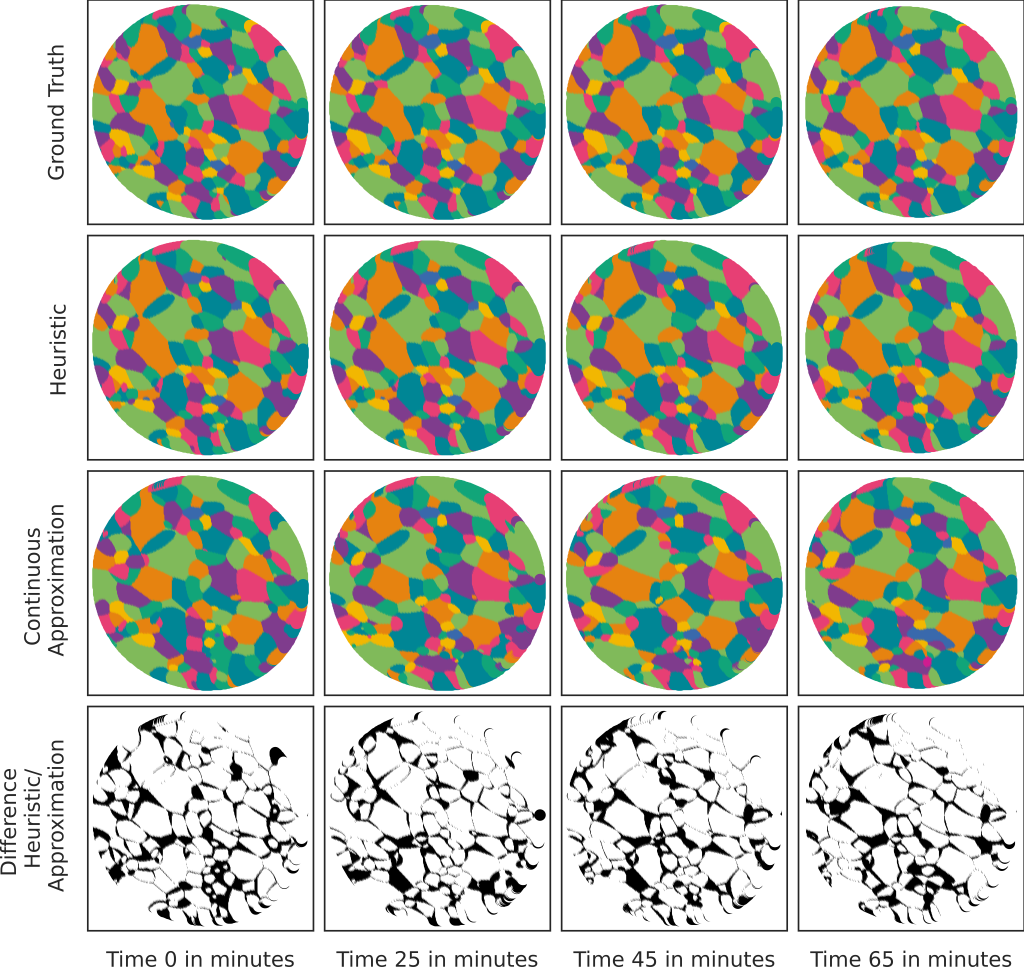}
    \caption{Visual comparison of diagrams obtained from full data (available for $t_0,\dots,t_{14}$) and the interpolation/extrapolation model that uses all data except for the respective moment in time. A fixed 2D slice through the 3D data set at times $t_0,$ $t_5,$ $t_9$ and $t_{13}$ (top row);  the corresponding diagram representation obtained via H1 (2nd row) and, respectively, via the interpolation/extrapolation model that uses only data from the other moments in time and best-fitting quadratic polynomials (3rd row); symmetric difference between the images from the 2nd and 3rd row (bottom row).}
    \label{fig:simulation}
\end{figure}

\Cref{fig:simulation} show a fixed 2D slice of the 3D polycrystal at various moments in time and the diagrams resulting from our model (in the cross-validation scheme described above). 

\subsection*{Size variation}\label{sec:variance}

In the remaining part of this section, we study how the growth process affects the size parameters $\apdweights_t$. As  \cref{fig:accuracy_simulation} shows, the accuracy of the model increases with the duration of the process. This may be seen as an indication that the employed diagram structures are energetically favorable. Complementing the analysis of \cref{sec:evaluation_H1-H2}, we will now empirically study how the size parameters $\apdweights$ of the diagrams change as the grains grow. 

Let us begin by recalling the interpretation for the similarities and differences of H1 and H2. In fact, based on the analysis of \cref{sec:difference} we suspected that for persistent grains the variances of the powers  $\Gamma^{\nicefrac{3}{2}}_t$ of the sizes chosen by H1 should be significantly smaller than for non persistent grains. As Tables \ref{tb:variance-persitent} and \ref{tb:variance-non-persitent} show, these variances are indeed all small but significantly smaller for the persistent grains. This means, on the one hand, that the sites do not vary much, and we might as well have employed the simpler heuristic~H2. On the other hand, in terms of the sites, non persistent grains are less \enquote{regular} than persistent grains. 

\begin{table}[H]
\begin{center}
\begin{tabular}{@{}l|ccccc@{}}
\toprule
& $t_{\ell+0}$      & $t_{\ell+1}$      & $t_{\ell+2}$      & $t_{\ell+3}$      & $t_{\ell+4}$    \\ \midrule
$\ell=0$ & $6.409\cdot 10^{-8}$ & $6.214\cdot 10^{-8}$ & $5.130\cdot 10^{-8}$ & $3.016\cdot 10^{-6}$ & $2.544\cdot 10^{-7}$  \\ 
$\ell=5$ & $6.719\cdot 10^{-6}$ & $5.839\cdot 10^{-6}$ & $1.057\cdot 10^{-5}$ & $7.025\cdot 10^{-4}$ & $9.815\cdot 10^{-7}$ \\ 
$\ell=10$ &  $2.286\cdot 10^{-4}$ & $7.425\cdot 10^{-5}$ & $5.749\cdot 10^{-5}$ & $1.185\cdot 10^{-3}$ & $1.823\cdot 10^{-5}$ \\ \bottomrule
\end{tabular}%
\caption{Variance of $\Gamma^{\nicefrac{3}{2}}_{t_\ell}$ for $\ell \in \{0,1,\ldots,14\}$ of persistent grains.}
\label{tb:variance-persitent}
\end{center}
\end{table}

\begin{table}[H]
\begin{center}
\begin{tabular}{@{}l|ccccc@{}}
\toprule
& $t_{\ell+0}$      & $t_{\ell+1}$      & $t_{\ell+2}$      & $t_{\ell+3}$      & $t_{\ell+4}$    \\ \midrule
$\ell=0$ & $7.570\cdot 10^{-3}$ & $1.040\cdot 10^{-2}$ & $4.915\cdot 10^{-3}$ & $1.974\cdot 10^{-2}$ & $2.993\cdot 10^{-3}$  \\ 
$\ell=5$ & $1.278\cdot 10^{-2}$ & $1.834\cdot 10^{-2}$ & $2.790\cdot 10^{-2}$ & $4.158\cdot 10^{-3}$ & $8.959\cdot 10^{-3}$ \\ 
$\ell=10$ &  $2.608\cdot 10^{-2}$ & $2.804\cdot 10^{-2}$ & $2.955\cdot 10^{-2}$ & $3.541\textbf{}\cdot 10^{-3}$ & $-$ \\ \bottomrule
\end{tabular}%
\caption{Variance of $\Gamma^{\nicefrac{3}{2}}_{t_\ell}$ for $\ell \in \{0,1,\ldots,14\}$ of non persistent grains.}
\label{tb:variance-non-persitent}
\end{center}
\end{table}

A closer look at \cref{tb:variance-persitent} shows that, somewhat unexpectedly, not even for persistent grains the variance of $\Gamma^{\nicefrac{3}{2}}_{t}$ decreases monotonously in time. One may have anticipated that with decreasing \enquote{system energy} the behavior of the volumes of the unit balls $\mathbb{B}^3_{A_i}(t)$ with respect to $A_i(t)=\bigl(\Sigma(G_i(t))\bigr)^{-1}$ become increasingly similar to that of the grain volumes $\nu_i(t)$. The results in \cref{tb:variance-persitent}, however, seem to suggest that, in the presence of grain disappearance, the neighboring grains fill the newly available space more directly and it requires additional energy to reach a more regular mass distribution.

\subsection*{Grain disappearance}\label{sec:grain-death}

Next, we use a standard logistic regression model for predicting the probability $p_i(t)$ of the binary response variable which signifies whether grain $i$ vanishes (response value 1) at the next observed moment $t+1$ or not (response value 0). As predictor variables we use the grain's volume and its size parameter as computed by H1. Hence, we estimate three parameters, factors $\beta_1$ and $\beta_2$ for each of the two variables and one parameter $\beta_0$  for the intercept, i.e., we use the regression model
\begin{equation*}
p_i(t)=  \frac{1}{1+e^{-(\beta_0+ \beta_1 \nu_i(t) +  \beta_2 \gamma_i(t))}}.
\end{equation*}
Note that the parameters $\beta_0,\beta_1$ and $\beta_2$ are independent of $t$. Hence the model is identical for all time steps in the sense that it depends only on the grain characteristic. Our \emph{null hypothesis} is that the size and volume parameter do not correlate with grain disappearance. As a significance level we choose $0.05$.

\Cref{tb:logistic_regression} depicts the computed coefficients and $p$-values. 
Let us, in addition, point out that the McFadden's Pseudo $R^2$ value is $0.322$, indicating an excellent model fit. 

\begin{table}[H]
\centering
\begin{tabular}{@{}lll@{}}
\toprule
          & coefficient & $p$-value          \\ \midrule
$\beta_0$ (intercept) & -4.0537     & \textless 0.0001 \\
$\beta_1$ (volume) & -0.0010     & \textless 0.0001 \\
$\beta_2$ (size parameter)  & -1.0924     & \textless 0.0001 \\ \bottomrule
\end{tabular}%
\caption{Coefficients estimated by maximum likelihood estimation and their $p$-values for the described logistic regression model.}
\label{tb:logistic_regression}
\end{table}
As it turns out, we can reject the null hypothesis at a significance level of 0.05. This suggests that the diagram size parameter $\apdweights_t$ is statistically significant in this model for predicting grain disappearance. 

\section{Final remarks}\label{sec:conclusion}

As we have seen, measurements at discrete times of a grain growth process can be turned into a continuous description of the process using diagram representations. Our findings about the increasing fit of anisotropic diagrams, the variance of $\Gamma^{\nicefrac{3}{2}}_{t}$ and the relevance of the sizes for predicting grain disappearance indicate that, in the course of the growth process, the grain scans tend towards particularly \enquote{simple} diagrams with essentially uniform size parameters. 

Since all computations are fast, and can, if necessary, generally even be speeded up by using H2 rather than H1, we believe that the described model has the potential of becoming a practical tool for simulation and further studies of grain growth in practice. A next step is to evaluate the model more extensively on various types of grain scan data. It would be desirable if a consolidating foundation of the model can be achieved based on first principles.

Finally, let us point out that the model allows for some flexibility to incorporate additional insight. It is possible, for instance, to incorporate available characteristics of grain growth into the curve fitting process. 

\section*{Acknowledgements}
We thank Henning Friis Poulsen for fruitful discussions and valuable comments on an earlier version of the present paper. We are also grateful to Henning Friis Poulsen and Allan Lyckegaard for providing the real-world data set used in this paper.

\printbibliography

@incollection{Gritzmann2017,
author = {Gritzmann, Peter and Klee, Victor},
booktitle = {CRC Handbook on Discrete and Computational Geometry, 3rd extended edition, (Eds. J.E. Goodman, J. O'Rourke and C.D. Toth)},
isbn = {9781315119601},
pages = {937--968},
title = {{Computational convexity}},
year = {2017}
}

@article{Altendorf2014,
abstract = {In the area of tessellation models, there is an intense activity to fully understand the classical models of Voronoi, Laguerre and Johnson-Mehl. Still, these models are all simulations of isotropic growth and are therefore limited to very simple and partly convex cell shapes. The here considered microstructure of martensitic steel has a much more complex and highly non convex cell shape, requiring new tessellation models. This paper presents a new approach for anisotropic tessellation models that resolve to the well-studied cases of Laguerre and Johnson-Mehl for spherical germs. Much better reconstructions can be achieved with these models and thus more realistic microstructure simulations can be produced for materials widely used in industry like martensitic and bainitic steels.},
author = {Altendorf, Hellen and Latourte, Felix and Jeulin, Dominique and Faessel, Matthieu and Saintoyant, Lucie},
doi = {10.5566/ias.v33.p121-130},
issn = {1854-5165},
journal = {Image Analysis {\&} Stereology},
%month = {may},
number = {2},
pages = {121},
title = {{3D reconstruction of a multiscale microstructure by anisotropic tessellation models}},
url = {https://www.ias-iss.org/ojs/IAS/article/view/1090},
volume = {33},
year = {2014}
}

@article{telley96b,
author = {Telley, Hubert and Liebling, Thomas M. and Mocellin, Alain},
doi = {10.1080/13642819608239125},
issn = {395--408},
journal = {Philosophical Magazine B},
number = {3},
pages = {409--427},
title = {{The Laguerre model of grain growth in two dimensions I. Cellular structures viewed as dynamical Laguerre tessellations}},
url = {https://doi.org/10.1080/13642819608239125},
volume = {73},
year = {1996}
}

@article{Brieden2017,
abstract = {The paper develops a general framework for constrained clustering which is based on the close connection of geometric clustering and diagrams. Various new structural and algorithmic results are proved (and known results generalized and unified) which show that the approach is computationally efficient and flexible enough to pursue various conflicting demands. The strength of the model is also demonstrated practically on real-world instances of the electoral district design problem where municipalities of a state have to be grouped into districts of nearly equal population while obeying certain politically motivated requirements.},
author = {Brieden, Andreas and Gritzmann, Peter and Klemm, Fabian},
doi = {10.1016/j.ejor.2017.04.018},
file = {:C$\backslash$:/Users/Maxim/Dropbox/Uni/Research/Literature/Brieden, Gritzmann, Klemm - 2017 - Constrained clustering via diagrams A unified theory and its application to electoral district design.pdf:pdf},
issn = {03772217},
journal = {European Journal of Operational Research},
keywords = {Combinatorial optimization,Constrained clustering,Electoral district design,Generalized Voronoi diagrams,OR in government},
%month = {nov},
number = {1},
pages = {18--34},
title = {{Constrained clustering via diagrams: A unified theory and its application to electoral district design}},
url = {https://linkinghub.elsevier.com/retrieve/pii/S037722171730351X},
volume = {263},
year = {2017}
}

@article{guilleminot11,
author = {Guilleminot, Johann and Noshadravan, Arash and Soize, Christian and Ghanem, Roger G.},
doi = {10.1016/j.cma.2011.01.016},
issn = {00457825},
journal = {Computer Methods in Applied Mechanics and Engineering},
%month = {apr},
number = {17-20},
pages = {1637--1648},
title = {{A probabilistic model for bounded elasticity tensor random fields with application to polycrystalline microstructures}},
url = {https://linkinghub.elsevier.com/retrieve/pii/S0045782511000314},
volume = {200},
year = {2011}
}

@article{sedivy16,
author = {{\v{S}}ediv{\'{y}}, Ondrej and Brereton, Tim and Westhoff, Daniel and Pol{\'{i}}vka, Leo{\v{s}} and Bene{\v{s}}, Viktor and Schmidt, Volker and J{\"{a}}ger, Ale{\v{s}}},
doi = {10.1080/14786435.2016.1183829},
issn = {1478-6435},
journal = {Philosophical Magazine},
%month = {jun},
number = {18},
pages = {1926--1949},
title = {{3D reconstruction of grains in polycrystalline materials using a tessellation model with curved grain boundaries}},
url = {https://www.tandfonline.com/doi/full/10.1080/14786435.2016.1183829},
volume = {96},
year = {2016}
}

@article{Lachenbruch1979,
author = {Lachenbruch, Peter A. and Goldstein, Matthew.},
doi = {10.2307/2529937},
issn = {0006341X},
journal = {Biometrics},
%%month = {mar},
number = {1},
pages = {69},
title = {{Discriminant Analysis}},
url = {https://www.jstor.org/stable/2529937?origin=crossref},
volume = {35},
year = {1979}
}

@article{Xu2009,
author = {Xu, Tao and Li, Mo},
doi = {10.1080/14786430802647065},
issn = {1478-6435},
journal = {Philosophical Magazine},
%month = {feb},
number = {4},
pages = {349--374},
title = {{Topological and statistical properties of a constrained Voronoi tessellation}},
url = {http://www.tandfonline.com/doi/abs/10.1080/14786430802647065},
volume = {89},
year = {2009}
}

@article{Lyckegaard2011,
abstract = {Accurate descriptions of 3D grain structures in polycrystalline materials are of key interest as the grain structure is closely correlated to the macroscopic properties of the material. In the present study, we investigate the accuracy of using Laguerre tessellations to represent 3D grain structures from only the spatial center of mass location and the volume of the grains. The ability of Laguerre tessellations to describe accurate grain shapes and topologies of real 3D grain structures are revealed by direct comparison to 3D reconstructions of an un-deformed meta-stable $\beta$ -titanium alloy obtained by phase-contrast micro-tomography. This study reveals that (volume weighted) Laguerre tessellations are superior to classical Voronoi tessellations when it comes to providing accurate representations of real 3D grain structures. Furthermore, although the Laguerre tessellations were only able to correctly describe the local arrangements of grains (i.e., the grain neighbors and number of grain facets) for 31.8{\%} of the investigated grains, the Laguerre tessellations were able to accurately describe statistical grain characteristics such as grain size distributions and grain neighbor distributions. {\textcopyright} 2011 WILEY-VCH Verlag GmbH {\&} Co. KGaA, Weinheim.},
author = {Lyckegaard, Allan and Lauridsen, Erik Mejdal and Ludwig, Wolfgang and Fonda, Richard Warren and Poulsen, Henning Friis},
doi = {10.1002/adem.201000258},
file = {:C$\backslash$:/Users/Maxim/Dropbox/Uni/Research/Literature/Lyckegaard et al. - 2011 - On the Use of Laguerre Tessellations for Representations of 3D Grain Structures.pdf:pdf},
issn = {14381656},
journal = {Advanced Engineering Materials},
%month = {mar},
number = {3},
pages = {165--170},
publisher = {Wiley-VCH Verlag},
title = {{On the use of Laguerre tessellations for representations of 3D grain structures}},
url = {http://doi.wiley.com/10.1002/adem.201000258 https://onlinelibrary.wiley.com/doi/10.1002/adem.201000258},
volume = {13},
year = {2011}
}

@article{Zhang2018,
author = {Zhang, Jin and Zhang, Yubin and Ludwig, Wolfgang and Rowenhorst, David and Voorhees, Peter W. and Poulsen, Henning F.},
doi = {10.1016/j.actamat.2018.06.021},
issn = {13596454},
journal = {Acta Materialia},
keywords = {Diffraction contrast tomography (DCT),Ferrite,Microstructure,Temporal evolution,X-ray synchrotron radiation},
%month = {sep},
pages = {76--85},
publisher = {Acta Materialia Inc},
title = {{Three-dimensional grain growth in pure iron. Part I. {S}tatistics on the grain level}},
url = {https://linkinghub.elsevier.com/retrieve/pii/S1359645418304890},
volume = {156},
year = {2018}
}

@article{barker16,
author = {Barker, James and Bollerhey, Gregor and Hamaekers, Jan},
journal = {Computational Materials Science},
pages = {54--63},
title = {{A multilevel approach to the evolutionary generation of polycrystalline structures}},
volume = {114},
year = {2016}
}

@article{Alpers2015,
abstract = {Characterizing the grain structure of polycrystalline material is an important task in material science. The present paper introduces the concept of generalized balanced power diagrams as a concise alternative to voxelated mappings. Here, each grain is represented by (measured approximations of) its centre of mass position, its volume and, if available, and by its second-order moments (in the non-equiaxed case). Such parameters may be obtained from 3D X-ray diffraction. As the exact global optimum of our model results from the solution of a suitable linear programme it can be computed quite efficiently. Based on verified real-world measurements, we show that from the few parameters per grain (3, respectively, 6 in 2D and 4, respectively, 10 in 3D) we obtain excellent representations of both equiaxed and non-equiaxed structures. Hence our approach seems to capture the physical principles governing the forming of such polycrystals in the underlying process quite well.},
author = {Alpers, Andreas and Brieden, Andreas and Gritzmann, Peter and Lyckegaard, Allan and Poulsen, Henning Friis},
doi = {10.1080/14786435.2015.1015469},
file = {:C$\backslash$:/Users/Maxim/Dropbox/Uni/Research/Literature/Alpers et al. - 2015 - Generalized balanced power diagrams for 3D representations of polycrystals.pdf:pdf},
issn = {1478-6435},
journal = {Philosophical Magazine},
keywords = {generalized balanced power diagrams,grains,linear programming,polycrystals,power diagrams,tessellations},
%month = {mar},
number = {9},
pages = {1016--1028},
publisher = {Taylor and Francis Ltd.},
title = {{Generalized balanced power diagrams for 3D representations of polycrystals}},
url = {http://www.tandfonline.com/doi/abs/10.1080/14786435.2015.1015469},
volume = {95},
year = {2015}
}

@article{teferra18,
author = {Teferra, Kirubel and Rowenhorst, David J.},
doi = {10.1080/09500839.2018.1472399},
file = {:C$\backslash$:/Users/Maxim/Dropbox/Uni/Research/Literature//Teferra, Rowenhorts, and Rowenhorst - 2018 - Direct parameter estimation for generalised balanced power diagrams.pdf:pdf},
issn = {0950-0839},
journal = {Philosophical Magazine Letters},
%month = {feb},
number = {2},
pages = {79--87},
title = {{Direct parameter estimation for generalised balanced power diagrams}},
url = {https://www.tandfonline.com/doi/full/10.1080/09500839.2018.1472399},
volume = {98},
year = {2018}
}

@unpublished{AFGK21b,
author = {Alpers, Andreas and Fiedler, Maximilian and Gritzmann, Peter and Klemm, Fabian},
title = {Turning grain maps into diagrams},
note={submitted},
year = {2022}
}

@article{Telley1992,
author = {Telley, Hubert and Liebling, Thomas M. and Mocellin, Alain and Righetti, Franco},
doi = {10.4028/www.scientific.net/MSF.94-96.301},
issn = {1662-9752},
journal = {Materials Science Forum},
%month = {jan},
number = {1},
pages = {301--306},
title = {{Simulating and modelling grain growth as the motion of a weighted Voronoi diagram}},
url = {https://www.scientific.net/MSF.94-96.301},
volume = {94-96},
year = {1992}
}

@article{kuehn08,
author = {K{\"{u}}hn, Martin and Steinhauser, Martin O.},
doi = {10.1063/1.2959733},
issn = {0003-6951},
journal = {Applied Physics Letters},
%month = {jul},
number = {3},
pages = {034102},
title = {{Modeling and simulation of microstructures using power diagrams: Proof of the concept}},
url = {http://aip.scitation.org/doi/10.1063/1.2959733},
volume = {93},
year = {2008}
}

@article{sedivy17,
abstract = {Parametric tessellation models are often used to approximate complex grain morphologies of polycrystalline microstructures. A big advantage of such models is the substantial reduction in disk space required to store large, three-dimensional data sets, especially when compared with voxel-based alternatives. By selection of an appropriate tessellation model, a reasonably small loss of information on the real grain shapes can usually be achieved. Special attention has recently been devoted to models based on ellipsoidal approximations fitted to each grain. Faces of these tessellations are portions of quadric surfaces whose parameters can be derived easily. In this paper, we deal with geometric features of the structure, notably curvatures and dihedral angles, which are closely related to the kinetics of grain growth. These characteristics are computed for ellipsoidbased tessellations fitted to two different aluminum alloys with nominal composition Al-3 wt{\%} Mg-0.2 wt{\%} Sc and Al-1 wt{\%} Mg. The results are then compared with estimations based on meshed empirical data. We observe that the model offers more consistent estimations of grain shape characteristics than do the meshed empirical data. Precise description of grain boundaries by the model is also promising with respect to possible applications of these tessellations in stochastic space-time modeling of grain growth.},
author = {{\v{S}}ediv{\'{y}}, Ondrej and Dake, Jules Mullen and {Krill III}, Carl Emil and Schmidt, Volker and J{\"{a}}ger, Ale{\v{s}}},
doi = {10.5566/ias.1656},
issn = {1854-5165},
journal = {Image Analysis {\&} Stereology},
%month = {mar},
number = {1},
pages = {5},
title = {{Description of the 3D morphology of grain boundaries in aluminium alloys using tessellation models generated by ellipsoids}},
url = {https://www.ias-iss.org/ojs/IAS/article/view/1656},
volume = {36},
year = {2017}
}

@article{Teferra2015,
author = {Teferra, Kirubel and Graham-Brady, Lori},
doi = {10.1016/j.commatsci.2015.02.006},
issn = {09270256},
journal = {Computational Materials Science},
%month = {may},
pages = {57--67},
title = {{Tessellation growth models for polycrystalline microstructures}},
url = {https://linkinghub.elsevier.com/retrieve/pii/S0927025615000671},
volume = {102},
year = {2015}
}

@article{Zhang2020,
title = {Grain boundary mobilities in polycrystals},
journal = {Acta Materialia},
volume = {191},
pages = {211-220},
year = {2020},
issn = {1359-6454},
doi = {10.1016/j.actamat.2020.03.044},
url = {https://www.sciencedirect.com/science/article/pii/S1359645420302317},
author = {Jin Zhang and Wolfgang Ludwig and Yubin Zhang and Hans Henrik B. S{\o}rensen and David J. Rowenhorst and Akinori Yamanaka and Peter W. Voorhees and Henning F. Poulsen}
}

@article{model0,
author={Cyril Stanley Smith}, 
title={Grain shapes and other metallurgical applications of topology},
journal={Metal Interfaces},
pages={65--133},
year={1952},
}

@article{model1,
author={John von Neumann}, 
title={Discussion: Shape of metal grains},
journal={Metal Interfaces},
pages={108--110},
year={1952},
}

@article{model2,
author={William W. Mullins}, 
title={Two-dimensional motion of idealized grain boundaries},
journal={Journal of Applied Physics},
volume={27},
pages={900--904},
year={1956},
}

@article{model3,
author={Robert D. MacPherson and David J. Srolovitz}, 
title={The von {N}eumann relation generalized to coarsening of three-dimensional microstructures},
journal={Nature},
volume={446},
pages={1053--1055},
year={2007},
}

@article{model5,
author={Hillert, Mats}, 
title={On the theory of normal and abnormal grain growth},
journal={Acta Metallurgica},
volume={13},
pages={227--238},
year={1965},
}

@article{graingrowthchallenges1,
author={Rios, Paulo Rangel and Z{\"o}llner, Dana}, 
title={Critical assessment 30: {G}rain growth - Unresolved issues},
journal={Material Science and Technology},
volume={34},
pages={629--638},
year={2018},
}

@article{graingrowthchallenges2,
author={Han, Jian and Thomas, Spencer L. and Srolovitz, David J.}, 
title={Grain-boundary kinetics: {A} unified approach},
journal={Progress in Material Science},
volume={98},
pages={386--476},
year={2018},
}

@article{modelnew,
author={Hu, Jianfeng and Wang, Xianhao and Zhang, Junzhan and Luo, Jun and Zhang, Zhijun and Shen, Zhijian}, 
title={A general mechanism of grain growth -{I}. {T}heory},
journal={Journal of Materiomics},
volume={7},
pages={1007--1013},
year={2021},
}

@book{graingrowthbook1,
author={Rollett, Anthony  and Rohrer, Gregory S. and Humphreys, John},
title={Recrystallization and Related Annealing Phenomena},
isbn={978-0-08-098235-9},
edition={3rd edition},
publisher={Elsevier, Amsterdam},
year={2017},
}

@book{poulsenbook,
author={Poulsen, Henning Friis},
title={Three-Dimensional {X}-Ray Diffraction Microscopy: Mapping Polycrystals and their Dynamics},
publisher={Springer, Berlin},
year={2004},
doi={10.1007/b97884},
isbn={978-3-540-22330-6},
}

@incollection{intcrystbook,
author={Poulsen, Henning Friis and Vaughan, Gavin B. M.},
editors={Gilmore, C. J. and Kaduk, J. A. and Schenk, H.},
booktitle={International Tables for Crystallography: Powder Diffraction},
title={Multigrain crystallography and three-dimensional grain mapping},
publisher={International Union of Crystallography, Hoboken},
volume={H},
year={2019},
isbn={978-1-118-41628-0},
}

@article{Spettl2016,
  title={Fitting {L}aguerre tessellation approximations to tomographic image data},
  author={Spettl, Aaron and Brereton, Tim and Duan, Qibin and Werz, Thomas and Krill III, Carl Emil and Kroese, Dirk P and Schmidt, Volker},
  journal={Philosophical Magazine},
  volume={96},
  number={2},
  pages={166--189},
  year={2016},
}

@article{fastLaguerre,
author = {Bourne, David P. and Kok, Piet J. J. and Roper, Steven M. and Spanjer, Wil D. T.},
doi = {10.1080/14786435.2020.1790053},
journal = {Philosophical Magazine},
number = {21},
pages = {2677--2707},
title = {Laguerre tessellations and polycrystalline microstructures: a fast algorithm for generating grains of given volumes},
url = {https://doi.org/10.1080/14786435.2020.1790053},
volume = {100},
year = {2020}
}

@book{learningbook,
author={Trevor Hastie and Robert Tibshirani and  Jerome Friedman},
title={The Elements of Statistical Learning: Data Mining, Inference, and Prediction},
edition={2nd edition},
publisher={Springer, New York},
year={2009},
doi={10.1007/978-0-387-84858-7},
isbn={978-0-387-84857-0},
}

@article{curvedgrainboundaries,
  author={Petrich, Lukas and Furat, Orkun and Wang, Mingyan and Krill III, Carl E., and Schmidt, Volker},
  title={Efficient fitting of {3D} tessellations to curved polycrystalline grain boundaries},
  journal={Frontiers in Materials},
  volume={8},
  pages={760602},
  year={2021},
}


	
\end{document}